# High-fidelity simulations of CdTe vapor deposition from a new bond-order potential-based molecular dynamics method

X. W. Zhou[1], D. Ward[2], B. M. Wong[3], F. P. Doty[2], J. A. Zimmerman[1], G. N. Nielson[4], J. L. Cruz-Campa[4], V. P. Gupta[5], J. E. Granata[6], J. J. Chavez[7], and D. Zubia[7]

[1] *Mechanics of Materials Department, Sandia National Laboratories, Livermore, CA 94550, USA*

[2] *Radiation and Nuclear Detection Materials and Analysis Department, Sandia National Laboratories, Livermore, CA 94550, USA*

[3] *Materials Chemistry Department, Sandia National Laboratories, Livermore, CA 94550, USA*

[4] *Advanced MEMS Department, Sandia National Laboratories, Albuquerque, NM 87185, USA*

[5] *Materials, Devices and Energy Technology Department, Sandia National Laboratories, Albuquerque, NM 87185, USA*

[6] *Photovoltaics and Grid Integration Department, Sandia National Laboratories, Albuquerque, NM 87185, USA*

[7] *Department of Electrical Engineering, University of Texas at El Paso, El Paso, TX 79968, USA*

***ABSTRACT***

CdTe has been a special semiconductor for constructing the lowest-cost solar cells and the CdTe-based $Cd_{1-x}Zn_xTe$ alloy has been the leading semiconductor for radiation detection applications. The performance currently achieved for the materials, however, is still far below the theoretical expectations. This is because the property-limiting nanoscale defects that are easily formed during the growth of CdTe crystals are difficult to explore in experiments. Here we demonstrate the capability of a bond order potential-based molecular dynamics method for predicting the crystalline growth of CdTe films during vapor deposition simulations. Such a method may begin to enable defects generated during vapor deposition of CdTe crystals to be accurately explored.

I. INTRODUCTION

The CdTe semiconductor compound is attractive for two important applications. In one application, CdTe films are used to produce terrestrial solar cells [1,2] at a production cost lower than any other photovoltaic technologies [3,4] presently available. This is because the material has good manufacturability, high solar energy absorption coefficients, and optimal band gaps for photoelectric conversion under solar radiation [5,6,7]. In the other application, the CdTe-based $Cd_{1-x}Zn_xTe$ (CZT) alloys have long been the dominant semiconducting material utilized in radiation detectors [8,9,10,11]. This is because these alloys have high atomic numbers for efficient radiation-atomic interactions, and ideal band gaps for both high electron-hole creation and low leakage current. Despite of the successful application of CdTe and CZT, the potential for further material improvement is still tremendous. For instance, the record energy conversion efficiency of CdTe solar cells achieved today is only ~16% as compared with the maximum theoretical prediction of 29% [5,12,13,14]. This sizeable discrepancy in efficiency is largely due to the various micro / nano scale defects formed within the multilayered films [4,7,12,15,16,17]. The property nonuniformity of the radiation detecting CZT crystals, on the other hand, has been the limiting factor for both a poor performance and a high material cost (arising from a low yield of usable portion of ingots) [8]. Grain boundaries and tellurium inclusions / precipitates are known to cause carrier transport nonuniformity [8,18,19]. Native defects such as vacancies, antisites, and interstitials can also affect properties [20,21]. The problem is that these defects are difficult to remove especially when their formation mechanisms are not understood. Note that some defects, such as dislocations [22], are prevalent in the materials but have not been well studied in the past. A high fidelity modeling approach that can reveal the formation of various defects as a function of processing conditions and ways to control them during growth is useful in guiding the growth optimization for further material improvement. This in turn impacts a wide range of applications including national security, medical imaging, environmental safety and remediation, and industrial processing monitoring.

Molecular dynamics (MD) simulations can in principle track the evolution of defects by integrating Newton's equations of motion for the positions of atoms subject to an interatomic potential. However, MD simulation of vapor deposition continuously remains to be a frontier research because it is extremely challenging to capture accurately the dynamics of a variety of atomic configurations statistically formed on a growth surface. In particular, if the interatomic potential used in the MD model overestimates the stability of a surface configuration, such a configuration may retain during the simulation to trigger unrealistically an amorphous growth. As a result, high-fidelity MD vapor deposition simulations of semiconductors are rarely possible due to the lack of an interatomic potential that is sufficiently accurate for a variety of local configurations. For example, past MD vapor deposition simulations of semiconductor [23,24,25,26] have used predominantly Stillinger-Weber potentials [27]. While Stillinger-Weber potentials can be readily parameterized to ensure the lowest energy tetrahedral crystal (e.g., diamond) and therefore the crystalline growth of such a crystal, they do not capture the property trends of other configurations encountered during growth [28]. As a result, these simulations are incapable of predicting accurate defect structures. Tersoff potentials [29], which can improve over Stillinger-Weber potentials in predicting property trends of different configurations, are difficult to parameterize to ensure the lowest energy for the equilibrium structure. As a result of poor parameterizations, many literature Tersoff potentials incorrectly predict amorphous growth [28].

In this work, we apply a new analytical bond order potential (BOP) [30] to perform a systematic study of vapor deposition of CdTe crystalline films, establishing qualitatively correlations between film quality (say, crystallinity, stoichiometry, and various antisite concentrations) and deposition conditions (mainly substrate temperature and vapor phase species ratio). Note that this BOP is analytically derived from quantum mechanical theories [31,32,33,34,35,36] and therefore further improves over Tersoff potentials on capturing property trends of different configurations. The illustration of crystalline growth simulations using the new BOP-based MD method means that defect formation mechanisms in CdTe/CZT crystals can now be studied at a fidelity level comparable to *ab initio* methods and a scale level on par with empirical molecular dynamics simulation methods. As a compelling example of significant importance, we demonstrate that our method can reveal the formation of misfit dislocations during the growth of a CdTe overlayer on a lattice-mismatched substrate.

## II. SIMULATION METHODS

**A. Interatomic potential**

Our previous study has indicated that there are two CdTe interatomic potentials available in the literature; one [37] is parameterized using the Stillinger-Weber potential format (SW) [27], and the other one [38] is parameterized using a Rockett modification [39] of the Tersoff potential format (TR) [29]. Unlike the Stillinger-Weber [27] and Tersoff/Brenner [29,40] types of potentials commonly used for semiconductor systems in the literature, the parameterization of the CdTe BOP [30] was performed iteratively by both testing the crystalline growth simulations and fitting the properties of a variety of clusters, lattices, surfaces, and defects. Details of the BOP are described previously [30]. For convenience, the global-, species-, pair-, and triple-parameters needed to fully define the BOP are listed respectively in Tables I-IV.

Table I. Global BOP parameters for CdTe.

| Symbol | --- |
|---|---|
| $\varsigma_1$ | 0.00001 |
| $\varsigma_2$ | 0.00001 |
| $\varsigma_3$ | 0.00100 |
| $\varsigma_4$ | 0.00001 |

Table II. Species-dependent BOP parameters for CdTe.

| Symbol | Cd | Te |
|---|---|---|
| $p_\pi$ | 0.420000 | 0.460686 |

Table III. Pair-dependent BOP parameters for CdTe.

| Symbol | Term | Cd-Cd | Te-Te | Cd-Te |
|---|---|---|---|---|
| $r_0$ | GSP reference radius (Å) | 3.1276 | 3.1626 | 3.1276 |
| $r_c$ | GSP characteristic radius (Å) | 3.1276 | 3.1626 | 3.1276 |
| $r_1$ | cutoff start radius (Å) | 3.7303 | 3.8046 | 4.0138 |
| $r_{cut}$ | cutoff radius (Å) | 4.3330 | 4.4465 | 4.9000 |
| $n_c$ | GSP decay exponent | 2.800000 | 2.799998 | 2.811251 |
| m | GSP attractive exponent | 3.263155 | 2.458846 | 2.587831 |
| n | GSP repulsive exponent | 1.553883 | 1.223306 | 1.287478 |
| $\phi_0$ | repulsive energy prefactor (eV) | 0.186369 | 0.876912 | 0.631440 |
| $\beta_{\sigma,0}$ | σ bond integral prefactor (eV) | 0.238318 | 0.782635 | 0.825290 |
| $\beta_{\pi,0}$ | π bond integral prefactor (eV) | 0.097599 | 0.531205 | 0.031743 |
| $c_\sigma$ | empirical $\Theta_\sigma$ parameter | 0.561130 | 1.014809 | 1.286955 |
| $f_\sigma$ | band-filling parameter | 0.431863 | 0.331227 | 0.5 |
| $k_\sigma$ | skewing prefactor | 15.00000 | -2.86019 | 0 |
| $c_\pi$ | empirical $\Theta_\pi$ parameter | 1 | 1 | 1 |
| $a_\pi$ | prefactor for $\Theta_\pi$ | 1 | 1 | 1 |

Table IV. Three-body-dependent BOP parameters for CdTe

| Symbol | jik | | | | | |
|---|---|---|---|---|---|---|
| | CdCdCd | TeTeTe | TeCdTe | CdCdTe | CdTeCd | CdTeTe |
| $p_\sigma$ | 1 | 1 | 1 | 1 | 1 | 1 |

| $b_\sigma$ | 0.762039 | 0.669623 | 0.200000 | 1.000000 | 0.2000000 | 0.999854 |
|---|---|---|---|---|---|---|
| $u_\sigma$ | -0.40000 | -0.14152 | -0.38336 | 0.099711 | -0.400000 | -0.00393 |

Detailed comparison of all three potentials with density functional theory (DFT) calculations has been documented [28,30]. Here we only highlight how the BOP improves over other models. As such, predictions from the three potentials on cohesive energies of a variety of elemental and compound phases and various defect energies for the equilibrium CdTe crystal are compared with DFT calculations in Fig. 1. Fig. 1(a) indicates that the cohesive energies calculated using the BOP model (the blue lines) are considerably closer to those predicted by DFT (the red lines) than the corresponding results of the SW and TR potentials. Most importantly, the BOP correctly specifies the lowest energies for the equilibrium phases for both elements and the compound (i.e., hcp Cd, A8 Te, and zb CdTe), with the calculated cohesive energies of the lowest energy phases matching the corresponding experimental values (unfilled stars). In sharp contrast, the lowest energy phases are calculated to be dc Cd, dc Te, and zb CdTe by the SW potential and dc Cd, bcc Te, and CsCl by the TR potential, where the only correct result is the zb CdTe structure by the SW potential. These results indicate that the TR potential cannot be used to study any of the equilibrium Cd, Te, and CdTe phases as the structures are not even stable in MD simulations. While the SW potential can be used in some sort of MD simulation to study the equilibrium CdTe phase, caution should be taken in explaining the results concerning defects as the potential is not transferrable to Cd and Te (and hence the defective) regimes.

Fig. 1(b) indicates that the energy trends (i.e., the sequence in which energies are ordered) of various defects calculated using the BOP method are in general much closer to the DFT calculations than the results from the SW and TR potentials. In particular, the SW potential indicates that the Cd antisite and Te antisite have the lowest energies (0.74 – 0.80 eV) and should be the dominant defects, whereas the DFT calculations indicate very high energies (2.12 – 3.71 eV) for these two defects. Even worse, the TR potential indicates that the Cd antisite has an extremely low energy of 0.18 eV, and at least 5 defects have a lower energy than Te vacancy which is the lowest energy defect in DFT calculations. Note that the defect energies are calculated using the lowest energy elemental phases as the reference states [41,42,43]. Consequently, the wrong values of defect energies from the SW and TR potentials are on top of the wrong elemental structures. On the other hand, both BOP and DFT give the lowest energy for Te vacancy. In fact, apart from a relatively more deviation of the energy of the <110> Cd dumbbell interstitial, BOP captures well the energies of the other three lowest energy (and hence the most important) defects in Te vacancy, Cd interstitial, and Cd at Te antisite.

The present study examines the growth of CdTe on a (010) surface. Detailed (010) energy phase diagrams predicted by the BOP have been evaluated [30]. The results indicated that the preferred (010) CdTe surfaces are Te-rich (2×1) and Cd-rich (1×1) (this surface can also be considered as bulk terminated). The Te-rich (2×1) surface reconstruction is in good agreement with experimental observation [44,45]. The Cd-rich (1×1) surface predicted by BOP appears to differ from the experimentally observed c(2×2) surface in the Cd-rich environments [44,46]. However, it should be noted that the Cd atoms do not form dimers on the (1×1) surface, which is in good agreement with experiments [46] but has not been predicted by other potentials or DFT [28]. Based on the discussions above, it is expected that BOP will produce more accurate results than the other methods currently available in literature.

Diffusion of single Cd atom on Te-terminated surface and single Te atom on Cd-terminated surface is examined using MD simulations at a high temperature of 1200 K. The surface atoms are seen to hop directly from one lattice site to another for both cases. Because the Cd surface does not form dimers, there is a single diffusion path associated with the Te atom diffusion on Cd surface. On the other hand, multiple diffusion paths are found for the Cd diffusion on Te surface due to different local surface reconstructions. For reference, we provide the activation energy barrier for Te diffusion on Cd surface of ~0.9 eV as calculated following the approach by Liu et al [47].

**B. Molecular dynamics model**

For the CdTe-on-CdTe vapor deposition simulations, an initial substrate of a zb crystal containing 216 Cd atoms and 216 Te atoms with 6 (101) layers in the x- direction, 12 (040) layers in the y- direction, and 6 ($\bar{1}$01) layers in the z- direction is first created using the equilibrium lattice constant. The top y surface is terminated by Cd initially. Periodic boundary conditions are used in the x- and z- directions so that the system can be viewed as infinitely large in these two directions. A free boundary condition is used in the y- direction to enable deposition on the top (040) surface. During the simulations, the bottom 3 (040) layers are fixed. A simulated growth temperature is created by assigning velocities to each of the remaining atoms based upon a Boltzmann probability distribution. The subsequent evolution of positions and velocities of the system atoms is then solved from interatomic forces and Newton's equations of motion using Nordsieck numerical integration algorithm [48]. A Lagrangian formalism that enables the periodic lengths to change during simulation [49] is adopted to relax stresses. To ensure a constant substrate temperature, all the atoms above the fixed region are controlled using a Nose-Hoover isothermal algorithm [50]. Growth is simulated by injecting Cd and Te vapor species from random locations far above the surface, and the injecting frequencies of the two species are determined from the simulated deposition rate and the Te:Cd vapor species ratio. The injected atoms are all assigned a remote incident kinetic energy. To capture the adatom incident energy effects, the newly added adatoms are not isothermally controlled until they are fully incorporated into the film, and their initial kinetic and potential (latent heat release) energies are fully dissipated. Simulations are performed at various substrate temperatures T between 700 and 1200 K, adatom incident energies $E_i$ between 0.1 and 1.0 eV, deposition rates R between 2 and 12 nm/ns, vapor species ratio Te:Cd between 0.8 and 1.5, two different tellurium vapor species of atomic Te and molecular $Te_2$ (cadmium vapor species is always in atomic Cd form), and a constant adatom incident angle θ = 0º (i.e., the adatom initial impact direction is normal to the surface).

For the simulation of a CdTe overlayer growth on a lattice mismatched substrate, a larger initial CdTe crystal containing 900 Cd atoms and 900 Te atoms with 25 (101) layers in the x- direction, 12 (040) layers in the y- direction, and 6 ($\bar{1}$01) layers in the z- direction is used. To simulate the lattice mismatch in the x- direction, the x- dimension of the initial crystal is stretched by 10%. Again the bottom 3 (040) layers of atoms are fixed during simulation. To prevent the expanded substrate from relaxing back, a constant volume algorithm is used in the MD simulation. Otherwise the simulation method is the same as described above. As a representative case, only one simulation is performed using an adatom incident energy of 0.1 eV, a stoichiometric Te:Cd vapor species ratio of 1 (molecular $Te_2$), a substrate temperature of 1000 K, and a deposition rate of ~2.4 nm/ns.

Due to the constraints of computational cost, MD simulations employ accelerated deposition rates. As will be discussed below, the accelerated deposition rates reduce the diffusion time of surface atoms before they are buried by subsequently deposited atoms. This problem can be mitigated using elevated substrate temperatures to accelerate the surface diffusion [51]. To provide a rough estimate of the effect, the simulated substrate temperature T at the simulated deposition rate R can be approximately converted to a temperature $T_x$ that corresponds to a different deposition rate $R_x$ by equating the diffusion distances of surface atoms obtained under the two different deposition rates [53]:

$$T_x = \frac{T}{\frac{k \cdot T}{Q} \ln\left(\frac{R}{R_x}\right) + 1}. \tag{1}$$

Here k is Boltzmann constant, and Q can be viewed as apparent activation energy of the structure-determining surface diffusion mechanism (i.e., the one that dictates the difference between the high and the low deposition rates). As will be discussed below, Eq. (1) can be to extract useful information when carefully applied.

## III. RESULTS

### A. Deposition rate effects

The deposition rates used in the MD simulations are orders of magnitude higher than those used in experiments. The main effects of accelerated deposition rates on film structures can be understood in the following: Consider that when adatoms are first condensed on a growth surface, they must quickly move to some nearby local minimum energy sites on the surface since the process is either associated with a small energy barrier or is barrierless. These local minimum energy sites are likely to be the lattice sites so that even these adatoms do not further move, they can still be buried by subsequently deposited atoms to result in a crystalline growth. This fast initial relaxation of adatoms to the nearby surface lattice sites can be captured by MD simulations at the accelerated deposition rates. Once they occupy the surface lattice sites, the adatoms can undergo various diffusion processes to migrate to lower energy sites that may be a further distance away, leading to the evolution of surface morphology towards more stable configurations. The diffusion jumps of surface atoms from one lattice site to another are longer thermally activated processes as they are associated with significant energy barriers. It can be seen that the net effects of the accelerated deposition rates used in MD simulations are to cause the surface atoms to be rapidly buried into the bulk so that their surface diffusion is significantly underestimated.

While the simulated deposition rates are much higher than those in reality, the overall trends between film structure and deposition rate obtained in simulations can still be useful. First, the underestimation of the surface diffusion at accelerated deposition rates may be mitigated by using an elevated substrate temperature to promote the diffusion. While this is a simple approximation, the simulated structure vs. deposition rate relation may more closely reflect the real structure vs. deposition rate trends. In addition, there exists a substrate temperature-dependent critical deposition rate above which amorphous films are always obtained regardless of the interatomic potential. This occurs because above this critical rate, the adatoms do not even have time to relax to the nearby lattice sites before being fully buried. With elevated temperatures, the critical deposition rate for crystalline to amorphous transition can be captured with the MD simulations. As will describe in the following, such simulations allow us to determine an apparent Gibbs free energy barrier associated with the transition between amorphous and crystalline films. This energy barrier can be used to understand experimental conditions (temperature and deposition rate) that control the formation of crystalline or amorphous films. Here we first examine the effect of the deposition rate.

*1. Atomic structure observation*

Simulations are performed at a constant substrate temperature of 1200 K, a constant adatom energy of 0.1 eV, a constant Te:Cd vapor species ratio of 1, and various deposition rates between 2 and 14 nm/ns, using the molecular tellurium vapor phase $Te_2$. The 0.1 eV adatom energy accurately models the thermal deposition processes commonly used to grow semiconductor films and the molecular tellurium vapor phase reflects well the experiments [54] as this phase is relatively stable. Examples of the simulated CdTe films obtained at ~11.2, ~8.1, ~5.4, and ~2.7 nm/ns are shown respectively in Figs. 2(a) – 2(d). In Fig. 2, the parameters $\xi$ and X measure crystallinity and stoichiometry as described in Appendix, and $f_{Cd@Te}$ and $f_{Te@Cd}$ represent Cd@Te and Te@Cd antisite concentrations respectively (for reference, $\xi = 1$, $X = 1$, $f_{Cd@Te} = 0$, and $f_{Te@Cd} = 0$ correspond to an ideal zb CdTe crystal at 0 K). It can be seen from Fig. 2(a) that at a high deposition rate of ~ 11 nm/ns, the atomic structure near the surface region is amorphous, and a portion of the film below the surface appears to be a metastable lattice different from the zb CdTe. Not surprisingly, Fig. 2(a) is accompanied by a low crystallinity parameter and high antisite defect concentrations as compared to Figs. 2(b)-2(d). This poor structure indicates that the high deposition rate of 11.2 nm/ns is likely to be above the critical deposition rate discussed above so that adatoms do not have sufficient time to relax to the nearby zb lattice sites before being buried by the new adatoms.

When the deposition rate is reduced to ~ 8 nm/ns as shown in Fig. 2(b), adatoms have more time to relax upon condensation, and hence the majority of the deposited film appears to be in the zb lattice. This is accompanied by an increase in the crystallinity parameter and a decrease in the antisite concentrations. When the deposition rate is further reduced to ~ 5 nm/ns in Fig. 2(c), adatoms further relax upon condensation. While the surface region is still not fully relaxed to the zb lattice, it is thinner than that seen in Fig. 2(b). This is consistent with a further increase in the crystallinity parameter and a decrease in the antisite defect concentrations.

Finally, when the deposition rate is reduced to ~ 3 nm/ns in Fig. 2(d), adatoms fully relax to the zb lattice sites upon condensation. Correspondingly, the film exhibits a highly crystalline zb structure. This configuration is also associated with the highest crystallinity parameter ($\xi$ ~ 0.8), and the lowest antisite defect concentrations. Note that $\xi = 1$ corresponds to an ideal zb crystal at a temperature of 0 K. Fig. 2(d) suggests that at a temperature of 1200 K, $\xi$ ~ 0.8 already corresponds to a highly crystalline crystal, and a deposition rate of ~ 3 nm/ns is likely lower than the critical deposition rate for crystalline growth of CdTe films.

*2. Crystallinity and stoichiometry trends*

The configurations shown in Fig. 2 give a panoramic view of the various atomic structures and allow a qualitative visual analysis of the film quality. To further explore the film structures, quantitative crystallinity ($\xi$) and stoichiometry (X) parameters are calculated as a function of deposition rate and the results are shown in Fig. 3. In Fig. 3 and the following figures (Figs. 4, 8, 9, 10, 12, 14), the lines are not fitted to the data but are used to guide the eye. The shaded region here and in Fig. 4 corresponds to a deposition rate range between 0.633 and 8.865 mm/s if the 1200 K elevated deposition temperature is converted to a common experimental growth temperature of 650 K using Eq. (1) and a surface diffusion barrier of $Q = 0.9$ eV. This conversion requires an underlying assumption that the structure-determining long time surface diffusion neglected by the accelerated deposition rates is primarily the single atom diffusion. Hence, the shaded region only serves as a reference as how the simulated results may correspond to experimental observations, and our analysis will still be based upon the simulated conditions. As can be seen from Fig. 3, reducing the deposition rate generally resulted in a continuous improvement of crystallinity within the deposition rate range between 2 and 14 nm/ns. In particular, a critical deposition rate of ~ 3 nm/ns corresponding to a high crystallinity parameter of $\xi \sim 0.8$ can be obtained from Fig. 3. On the other hand, Fig. 3 shows that the film stoichiometry is close to the ideal value of one for the entire deposition rate range explored, suggesting that the film stoichiometry may be primarily determined by the vapor species ratio (the vapor species ratio used here is Te:Cd = 1) and less sensitive to the deposition rate.

*3. Antisite concentration trends*

The fraction of Cd@Te and Te@Cd antisites with respect to the number of lattice sites in the deposited films is also calculated to quantify the deposition rate effect on defect incorporation. The results of this calculation are shown in Fig. 4. It can be seen that the antisite defect concentrations continuously decrease as the deposition rate is reduced. This arises from the antisites being associated with high energies and not being stable configurations. When the deposition rate is high, the high energy surface antisites do not have sufficient time to reconstruct to lower energy configurations before being buried. As a result, high antisite defect concentrations are obtained; when the deposition rate is reduced, these defects can reconstruct to lower energy zb configurations, resulting in the reduction of the antisite concentrations.

**B. Deposition species effects**

While CdTe films are conveniently grown from atomic Cd and molecular $Te_2$ vapor species [54], they can also be grown from all-atomic vapor species. To explore if the species of vapor phases can affect the film structures, simulations are performed using atomic forms of Cd and Te adatoms at a substrate temperature of 1200 K, a deposition rate near 2 nm/ns, a Te:Cd vapor species ratio of 1, and two adatom energies of 0.1 and 1.0 eV. The configurations obtained are shown respectively in Figs. 5(a) and 5(b) for the two energies. Comparing Fig. 5(a) and Fig. 2(a) indicates that the film structure obtained using the atomic tellurium species is relatively less crystalline. This is consistent with a smaller crystalline parameter ξ, a larger deviation of the stoichiometric parameter X from the ideal value of 1, and high antisite concentrations $f_{Cd@Te}$ and $f_{Te@Cd}$ in the atomic case than in the molecular case. Note that when a $Te_2$ molecule is condensed on a surface, it is readily dissociated into two Te atoms. This dissociation absorbs energy from the surface and two closely separated Te atoms also provide strong bonding to the nearby Cd atoms (the Cd-Te interaction is relatively strong). As a result, this reduces the re-evaporation of the volatile cadmium. On the other hand, an atomic Te adatom may promote the evaporation of cadmium. This can be verified from Fig. 5(a), where the stoichiometry parameter is noticeably larger than the ideal value of one (i.e., Cd depletion), and the overall deposition rate is also smaller than that in Fig. 2(a) (the adatom injection frequencies used in both cases are the same). Because the film obtained using the atomic Te vapor species is not in an ideal stoichiometry, it is not surprising that it is associated with a reduced crystallinity and increased antisite defect concentrations.

Comparing Fig. 5(b) with Fig. 5(a) indicates that increasing the adatom energy does not significantly change the film structure with the exception that Fig. 5(b) has a flatter surface than Fig. 5(a) due to impact-induced flattening effects [55,56]. Interestingly, Fig. 5(b) also shows the evaporation of some $Te_2$ molecules within the time scale of our simulations. It is important to point out that while only atomic Te vapor species are used in our simulations, the excess Te atoms on the surface can recombine to form molecules and re-evaporate into the vapor phase under the high energy conditions. This finding is in good agreement with the experimental observation that $Te_2$ readily forms in the vapor [54].

**C. Adatom energy effects**

While the adatom energy in the thermal evaporation deposition processes is usually limited to around 0.1 eV, many sputter deposition processes enable adatom energies to be maintained at hyper thermal energies of greater than 1.0 eV. Fig. 5 already explores the adatom energy effects using the atomic Te vapor species. To further examine the adatom energy effects, simulations are performed at a substrate temperature of 1200 K, a deposition rate near 2.7 nm/ns, a Te:Cd vapor species ratio of 1, a molecular $Te_2$ vapor species, and various adatom energies between 0.1 and 1.0 eV. The resulting atomic configurations obtained at two adatom energies of 0.6 and 1.0 eV are respectively shown in Figs. 6(a) and 6(b). Figs. 2(a), Figs. 6(a), and 6(b) indicate that the film atomic scale configurations, crystallinity $\xi$, stoichiometry X, and antisites fractions $f_{Cd@Te}$ and $f_{Te@Cd}$ obtained at different adatom energies are comparable. These results strongly suggest that the thermally-activated diffusion processes may be more important in controlling the structures of CdTe films than the short-time, adatom energy-induced impact processes.

**D. Deposition temperature effects**

Substrate temperature and the vapor phase species ratio (representing the chemical potential of the species in the vapor) are the two primary processing conditions for vapor deposition of semiconductor compounds. In the following discussion, we first explore the effect of substrate temperature on CdTe growth.

*1. Atomic structure observation*

Simulations are performed at a deposition rate near 3 nm/ns, a Te:Cd vapor species ratio of 1, a molecular $Te_2$ vapor species, an adatom energy of 0.1 eV, and various substrate temperatures between 500 and 1200 K. The atomic configurations at four selected substrate temperatures of 700, 900, 1000, and 1100 K are shown in Figs. 7(a) – 7(d) respectively. It can be seen that at a low substrate temperature of 700 K, the film is largely amorphous as shown in Fig. 7(a). This growth pattern emerges since the critical deposition rate for crystalline growth is below 3 nm/ns at 700 K. However, when the temperature is increased to 900 K in Fig. 7(b), a majority of the film becomes crystalline with the exception of a thin surface layer which still retains some irregular features. When the substrate temperature is further increased to 1000 K in Fig. 7(c), the irregular layer at the surface becomes thinner. Finally, when the substrate temperature is increased to 1100 K in Fig. 7(d), the entire film becomes highly crystalline. These observations verify that as the temperature is increased, adatoms become increasingly mobile and are therefore more likely to relax to the low-energy lattice sites on a short time scale. As a result, increasing the substrate temperature has a similar effect to decreasing the deposition rate as shown in Fig. 2.

*2. Crystallinity and stoichiometry trends*

To get a more quantitative picture of the temperature effects on film structure, crystallinity and stoichiometry parameters are calculated for all the substrate temperatures, and the results are shown in Fig. 8 with the filled diamonds and circles (the open diamonds will be described later). The shaded region here and in Figs. 9 and 10 corresponds to a temperature range between 282 and 421 K if the accelerated deposition rate of ~3 nm/ns is converted to a realistic rate of 0.3 μm/s using Eq. (1) and a surface diffusion barrier of $Q = 0.9$ eV. Again the purpose of this conversion is to provide a reference to the experimental conditions, and the analysis will be based upon the simulated conditions.

It can be seen that at the deposition rate of 3 nm/ns, the crystallinity is very poor ($\xi = 0.3$ or below) when the substrate temperature is below 700 K. As the temperature is increased from 700 K, the crystallinity first undergoes a sharp improvement and then reaches a saturated value around 0.8 at a substrate temperature of 900 K. This indicates that at the chosen simulated deposition rate of 3 nm/ns, MD simulations of vapor deposition of CdTe films must be carried out at substrate temperatures above 900 K in order to reveal any insights on crystalline growth. Fig. 8 also indicates that at low simulated substrate temperatures of ~ 500 K, the stoichiometry of the films is significantly below the ideal value of 1 at the stoichiometric vapor ratio Te:Cd = 1 (increasing the substrate temperature causes the film stoichiometry to approach this ideal value). The lower film stoichiometry at low substrate temperatures indicates that the sticking of the tellurium vapor species on the growth surface is lower than that of the cadmium species. Furthermore, the low stoichiometry (Te-depletion) of the films promotes the formation of defects such as Te vacancies, Cd@Te antisites, and Cd interstitials. In general, the sticking of tellurium and cadmium atoms can be both maximized if the growth of the zb CdTe film is ideally stoichiometric because this phase is the most stable. This implies that if the film is Cd-rich, the evaporation of Cd becomes more significant than that of Te; and likewise, if the film is Te-rich, the evaporation of Te becomes more significant than that of Cd. Increasing the substrate temperature promotes this process towards equilibrium and, therefore, results in more stoichiometric films.

As described above, the shaded regions in Figs. 3, 4, and 8-10 approximately relate the simulated conditions to experimental conditions using Eq. (1) with an underlying assumption that surface evolution during a low growth rate proceeds predominantly through single adatom diffusion towards nearby islands. It is recognized that the amorphous-to-crystalline transition observed in our simulations proceeds through adatom transition from its initially condensed site to nearby lattice sites. Multiple paths exist for these diffusions due to random adatom condensation. These paths can also be thermally activated, but their energy barriers must be small (or barrierless if an adatom directly hits a lattice site) as the process is seen to be very fast.

In general, multiple diffusion paths can be captured as the entropy effect so that the amorphous-to-crystalline transition exhibits an apparent Gibbs free energy barrier Q. This concept can also be tested using Eq. (1). First, it should be noted that the crystallinity values obtained at different deposition rates and a constant temperature of 1200 K (i.e., the filled diamonds in Fig. 3) can be converted to those at different temperatures and a constant deposition rate of around 3 nm/ns using Eq. (1) for any given value of Q. If Eq. (1) is valid for the amorphous-to-crystalline transition, the converted data should match the direct $\xi$ vs. T MD data (i.e., the filled diamonds in Fig. 8). By minimizing the deviation of the converted data from the curve drawn through the filled diamonds in Fig. 8, we obtain an apparent activation (Gibbs free) energy barrier of Q = 0.224 eV. The resulting converted $\xi$ vs. T data are shown in Fig. 8 using unfilled diamonds. It can be seen that other than the anomaly point which incorrectly shows a higher crystallinity at a lower growth temperature, the converted data agree remarkably well with the unconverted data. In particular, the converted data also exhibit a sharp change in crystallinity when the growth temperature changes from 700 to 900 K. These results suggest that Eq. (1) is a reasonable approximation to extrapolate data for a given thermally activated process. More importantly, the apparent activation energy barrier of 0.224 eV has important implications for understanding the experimental growth mechanisms especially the effects of the growth conditions on the transition from amorphous to crystalline films.

*3. Antisite concentration trends*

Next, the Cd@Te and Te@Cd antisite concentrations are also calculated, and the results are shown as a function of substrate temperature in Fig. 9. It can be seen that the two antisite concentrations both decrease as the substrate temperature is increased. This can be attributed to two mechanisms. First, the improvement in film stoichiometry due to the increase in substrate temperature naturally results in the reduction in the antisite concentrations. Second, the antisites are naturally associated with higher energies. The increase in substrate temperature increases the mobility of surface atoms to anneal out these higher energy defects before they are buried into the bulk part of the films.

*4. Sticking coefficient trends*

Finally, to understand interactions between vapor and the growth surface, the sticking probability (ratio of atoms eventually incorporated in the film to total injected atoms) of the deposited atoms is calculated as a function of substrate temperature in Fig. 10. As expected, it can be seen that the sticking probability is substantially reduced as the substrate temperature is increased. In particular, we find that the Te sticking probability is lower on a Te-rich surface than on a Cd-rich surface, and the Cd sticking probability is lower on a Cd-rich surface than on a Te-rich surface. Increasing the substrate temperature promotes this effect, and provides a mechanism for improving film stoichiometry and reducing defect concentrations.

**E. Vapor phase species ratio effects**

The vapor species ratio is another important processing condition in the vapor deposition of semiconductor compounds as it directly controls the stoichiometry of the films. By capturing energies of a variety of structures, BOP can predict the vapor species ratio effects (i.e., chemical vapor deposition) that cannot be revealed by SW potentials. In this section we explore vapor species ratio effects and the various environmental conditions which affect them.

*1. Atomic structure observation*

To investigate the effects of vapor species ratios, simulations are performed at a deposition rate near 3 nm/ns, two substrate temperatures of 1000 and 1200 K, a molecular $Te_2$ vapor species, an adatom energy of 0.1 eV, and various vapor species ratios between Te:Cd = 0.8 and Te:Cd = 1.5. Selected atomic configurations at two Te:Cd vapor species ratios of 0.8 and 1.2 and two substrate temperatures of 1000 and 1200 K are compared in Fig. 11. Fig. 11 indicates that at the lower temperature of 1000 K, the film obtained under the Cd-rich vapor condition (Te:Cd = 0.8) has a sharp crystalline structure, as shown in Fig. 11(a). This implies that the excess Cd atoms deposited on the surface can re-evaporate into the vapor at 1000 K. This behavior is not surprising since Cd is rather volatile with a low cohesive energy of -1.133 eV/atom [57]. On the other hand, Fig. 11(b) shows that the crystalline quality of the film becomes significantly reduced under Te-rich (Te:Cd = 1.2) vapor growth conditions at 1000 K. This indicates that the excess Te atoms deposited on the surface cannot be fully re-evaporated at 1000 K. Interestingly, Figs. 11(c) and 11(d) indicate that the crystalline quality is high at the higher temperature of 1200 K under both Cd-rich and Te-rich vapor growth conditions. This implies that at 1200 K, the re-evaporation of excess species on the growth surface is significant even for the less volatile Te, thereby promoting the formation of stoichiometric films.

*2. Crystallinity and stoichiometry trends*

The crystallinity and stoichiometry parameters are again calculated, and the results are shown in Fig. 12 as a function of the vapor phase Te:Cd ratio at two substrate temperatures of 1000 and 1200 K. It can be seen that at both temperatures, the crystallinity reaches a peak value of approximately 0.8 when the Te:Cd vapor species ratio is between 0.9 and 1.1. This is expected since the stoichiometric vapor ratio is most likely to result in a stoichiometric film composition required for the highly crystalline CdTe compound. Interestingly, the decrease in the film crystallinity is more significant under Te-rich vapor conditions than under Cd-rich conditions at the low temperature of 1000 K (this is due to the excess Cd being more likely to re-evaporate as described above). However, since the excess Te can easily re-evaporate at a high temperature, the difference between the Cd-rich and Te-rich vapor conditions becomes smaller at the high substrate temperature of 1200 K, as shown in Fig. 12.

As expected, Fig. 12 also indicates that an increase in the vapor phase Te:Cd ratio continuously results in an increase in the Te:Cd ratio in the deposited films, and the stoichiometric film composition occurs when the vapor Te:Cd ratio is close to (but slightly less than) unity. Furthermore, increasing the substrate temperature causes the film composition to be closer to the stoichiometric value at various vapor Te:Cd ratios. This is entirely consistent with the discussion above which explains that the high temperature induced evaporation helps adjust the film composition towards the stoichiometric value.

*3. Antisite concentration trends*

Cd@Te and Te@Cd antisite concentrations are calculated, and the results are shown in Fig. 13 as a function of the vapor Te:Cd ratio for the two substrate temperatures of 1000 and 1200 K. It can be seen that the two antisite concentrations both reach their minimum at a stoichiometric vapor Te:Cd ratio near 1. While the increase in the antisite concentrations due to a reduction in the vapor Te:Cd ratio is relatively insignificant, large antisite concentrations are obtained when the Te:Cd ratio is significantly larger than 1. Increasing temperature results in lower antisite concentrations. These observations are consistent with the discussion above that the film structure is more tolerant to the Cd-rich growth condition since the excess Cd atoms on the surface can readily re-evaporate, and the high temperatures promote stoichiometric films under Te-rich growth conditions since they enable the excess Te atoms on the surface to also re-evaporate.

*4. Sticking coefficient trends*

Finally, the sticking probability of the deposited atoms is calculated, and the results are plotted as a function of vapor Te:Cd ratio in Fig. 14 for the two substrate temperatures of 1000 and 1200 K. It can be seen that at the relatively low temperature of 1000 K, increasing the vapor Te:Cd ratio from 0.8 to 1.5 continuously increases the sticking probability. This steady increase occurs because the growth surface has excess Cd at the Cd-rich deposition condition of vapor Te:Cd ratio near 0.8. As a result, the volatile Cd easily re-evaporates, leading to a low sticking coefficient. However, as the vapor Te:Cd ratio increases, the surface becomes increasingly Te-rich, and since the Te atoms are less likely to re-evaporate, the sticking probability increases. A different situation arises at the relatively high substrate temperature of 1200 K where the sticking is seen to reach a maximum value at a vapor Te:Cd ratio near 1.2. At 1200 K, both excess Cd and excess Te atoms on the surface can re-evaporate. Only when the surface forms the lowest-energy stoichiometric CdTe compound, or the Te surface atoms are not too excessive, will the re-evaporation be low. This accounts for the overall trends observed in Fig. 14 and the other observations mentioned previously.

**F. Misfit strain effects**

Using the approach described in section II.B, a special case is studied where a CdTe overlayer is deposited on a lattice-mismatched substrate that is 10% larger. The growth simulation is performed using an adatom incident energy of 0.1 eV, a stoichiometric Te:Cd vapor species ratio of 1 (molecular $Te_2$), a substrate temperature of 1000 K, and a deposition rate of ~2.4 nm/ns. The atomic configurations obtained after 0.012 ns and 1.2 ns of deposition are shown respectively in Figs. 15(a) and 15(b), where the pink shade highlights the fixed region which can be viewed as the larger substrate, and the blue shade highlights the original (i.e., prior to deposition) CdTe layer on top of the larger substrate. Note that the original CdTe layer is perfect and completely epitaxial with the substrate as seen in Fig. 15(a). Fig. 15(b) indicates that after 1.2 ns of deposition, an edge-type misfit dislocation is formed near the interface as indicated by the red circle. This misfit dislocation exhibits two extra planes about 144.7$^o$ from the y- axis as indicated by the green lines, matching exactly the high-resolution transmission electron microscopic experiments [58]. Our BOP-based simulations are able to show that the misfit dislocation is first nucleated at the surface and then climbs vertically towards the interface. This discovery differs from the conventional mechanisms that misfit dislocations come from the pre-existing threading dislocations in the underlying substrate [59]. Note that unlike high energy defects such as vacancies, antisites, or interstitials which can be kinetically trapped by the accelerated deposition rate, the formation of misfit dislocations causes a reduction of system energy by releasing the misfit strain. This means that misfit dislocation formation is not an artifact of the accelerated deposition rate, and more misfit dislocations would occur should the deposition rate be reduced. As a result, the observation of misfit dislocations demonstrates a powerful ability of the BOP-based MD approach in exploring methods to reduce the property-limiting misfit dislocations for a variety of semiconductor multilayerd structures.

## IV. CONCLUSIONS

In this study, we have performed extensive MD simulations of CdTe vapor deposition using a new, first-principles-based BOP method. The successful application of the high-fidelity BOP formalism in MD simulations opens up a new opportunity to accurately predict defect formation mechanisms during crystal growth. The following conclusions have been reached :

(i) In order to have crystalline growth, the MD deposition rates need to be below ~ 3 nm/ns, and the substrate temperatures need to be above 900 K (cf. Fig. 8). This is determined by the adatom transition from an initial site of condensation to a nearby lattice site, and the process is associated with a small energy barrier of 0.224 eV. This energy barrier can be used to understand the experimental growth mechanisms especially the effects of the growth conditions on the transition from amorphous to crystalline films. It is also a useful benchmark value for explaining MD simulations.

(ii) If atomic Te vapor is used instead of the molecular $Te_2$ vapor, the vapor Te:Cd ratio needs to be reduced in order to create stoichiometric films. These findings are equivalent to keeping the tellurium chemical potential constant.

(iii) The film structure is not sensitively affected by the adatom energy.

(iv) A processing condition with vapor Te:Cd < 1 is more likely to produce stoichiometric CdTe films than a condition with vapor Te:Cd > 1 since excess Cd atoms are more likely to re-evaporate than excess Te atoms. A less stringent tolerance on the vapor species ratio can be obtained by increasing the substrate temperature where both excess Cd and excess Te atoms on the surface can re-evaporate.

(v) The BOP-based MD method can be used to examine detailed mechanisms of misfit dislocation formation in lattice-mismatched systems. As the energy efficiency of CdTe solar cells is currently limited by misfit dislocation formation which has been prohibitively difficult to examine in experiments, it would be extremely interesting to apply our techniques in large-scale MD simulations to study lattice-mismatch effects of semiconductor multilayers. We are currently performing further calculations to explore new nano structural designs to reduce misfit dislocations. With this in mind, we anticipate that BOP-based MD techniques will play a significant role in understanding and accurately predicting the material properties for many technologically-important systems.

**ACKNOWLEDGEMENTS**

This work is supported by the NNSA/DOE Office of Nonproliferation Research and Development, Proliferation Detection Program, Advanced Materials Portfolio, and The National Institute for Nano-Engineering (NINE). Sandia National Laboratories is a multi-program laboratory managed and operated by Sandia Corporation, a wholly owned subsidiary of Lockheed Martin Corporation, for the U.S. Department of Energy's National Nuclear Security Administration under contract DE-AC04-94AL85000.

**APPENDIX: CRYSTALLINITY AND STOICHIOMETRY PARAMETERS**

In this section, we describe a crystallinity parameter which is capable of quantifying the structural similarity of a deposited film with an ideal crystal. Suppose that a given adatom $i$ has a set of nearest neighbor atoms whose positions are notated as $\{\mathbf{r}_i\}$. If the center atom $i$ is assumed to be at a lattice site, then a set of nearest neighbor lattice sites can be determined based on the substrate orientation and lattice constants. The positions of these lattice sites which most closely match $\{\mathbf{r}_i\}$ are notated as $\{\mathbf{R}_i\}$. The mean square deviation of $\{\mathbf{r}_i\}$ from $\{\mathbf{R}_i\}$, $\Delta_i = \langle(\mathbf{R}_i - \mathbf{r}_i)^2\rangle$, represents the deviation of atom $i$'s environment from the crystalline configuration. A parameter, $\xi_i = \exp(-\alpha \cdot \Delta_i)$, which equals 1 at $\Delta_i = 0$ and continuously drops to zero when $\Delta_i$ is increased, can then be used to characterize the crystallinity of atom $i$. For convenience, we choose $\alpha = 1$. The crystallinity of the film can then be well-described by the average crystallinity of deposited atoms, $\xi = \sum_i \xi_i / N$, where N is the total number of the deposited atoms included in the summation. Based on these definitions, an ideal crystal at 0 K (i.e., no thermal oscillation) corresponds to a value of $\xi = 1$.

The ratio of Te to Cd atoms in the deposited films, $X = N_{Te}/N_{Cd}$, can be used to measure the stoichiometry of the film (this also reflects the concentrations of defects such as vacancies, antisites, and interstitials), where $N_{Te}$ and $N_{Cd}$ are respectively the total numbers of Te and Cd atoms in the deposited film.

**Figure Captions**

Fig. 1. Comparison of different models for (a) cohesive energies of a variety of Cd, Te, and CdTe phases; and (b) energies of various defects in the equilibrium zinc-blende CdTe crystal. Abbreviations: dc (diamond-cubic), sc (simple-cubic), bcc (body-centered-cubic), fcc (face-centered-cubic), hcp (hexagonal-close-packed), gra (graphite), grap (graphene), A8 (γ-Se), zb (zinc-blende), wz (wurtzite), B1 (NaCl), and B2 (CsCl), $V_{Cd}$ (Cd vacancy), $V_{Te}$ (Te vacancy), $Cd_{Te}$ (Cd at Te antisite), $Te_{Cd}$ (Te at Cd antisite), $Cd_i$ (Cd interstitial in Te tetrahedron), $Te_i$ (Te interstitial in Cd tetrahedron), $Cd_{i,<110>}$ and $Cd_{i,<100>}$ (Cd dumbbell interstitials along the <110> and <100> directions), and $Te_{i,<110>}$ and $Te_{i,<100>}$ (Te dumbbell interstitials along the <110> and <100> directions). Here the straight lines connecting the neighboring data points are used to guide the eye, and unfilled stars refer to the experimental values of the cohesive energies of the equilibrium phases [57].

Fig. 2. Film configurations obtained at a constant temperature of 1200 K and four different deposition rates of (a) ~11.2 nm/ns; (b) ~8.1 nm/ns; (c) ~5.4 nm/ns; and (d) ~2.7 nm/ns with diatomic $Te_2$ vapor. ξ: crystallinity; X: stoichiometry: $f_{Cd@Te}$ and $f_{Te@Cd}$: Cd@Te and Te@Cd antisite concentrations.

Fig. 3. Effect of deposition rate on film crystallinity and stoichiometry. The lines only serve to guide the eye. The shaded region corresponds to a converted deposition rate range of R between 0.633 and 8.865 mm/s using Eq. (1), a deposition temperature of T = 650 K, and a surface diffusion barrier of Q = 0.9 eV.

Fig. 4. Effect of deposition rate on film antisite concentrations. The lines only serve to guide the eye. The shaded region corresponds to a converted deposition rate range of R

between 0.633 and 8.865 mm/s using Eq. (1), a deposition temperature of T = 650 K, and a surface diffusion barrier of Q = 0.9 eV.

Fig. 5. Effect of tellurium vapor species (atomic Te instead of molecular $Te_2$) on film configuration at two adatom energies of (a) 0.1 eV and (b) 1.0 eV. $\xi$: crystallinity; X: stoichiometry: $f_{Cd@Te}$ and $f_{Te@Cd}$: Cd@Te and Te@Cd antisite concentrations.

Fig. 6. Film configurations obtained at a temperature of 1200 K, a deposition rate of ~2.7 nm/ns, and two different adatom energies of (a) 0.6 eV and (b) 1.0 eV with diatomic $Te_2$ vapor. $\xi$: crystallinity; X: stoichiometry: $f_{Cd@Te}$ and $f_{Te@Cd}$: Cd@Te and Te@Cd antisite concentrations.

Fig. 7. Film configurations obtained at a constant deposition rate near 3 nm/ns and four different temperatures of (a) 700 K; (b) 900 K; (c) 1000 K; and (d) 1100 K with diatomic Te2 vapor. $\xi$: crystallinity; X: stoichiometry: $f_{Cd@Te}$ and $f_{Te@Cd}$: Cd@Te and Te@Cd antisite concentrations.

Fig. 8. Effect of substrate temperature on film crystallinity and stoichiometry. The lines only serve to guide the eye. The shaded region corresponds to a converted deposition temperature range of T between 282 and 421 K using Eq. (1), a deposition rate of 0.3 μm/s, and a surface diffusion barrier of Q = 0.9 eV.

Fig. 9. Effect of substrate temperature on film antisite concentrations. The lines only serve to guide the eye. The shaded region corresponds to a converted deposition temperature range of T between 282 and 421 K using Eq. (1), a deposition rate of 0.3 μm/s, and a surface diffusion barrier of Q = 0.9 eV.

Fig. 10. Effect of substrate temperature on sticking coefficient. The lines only serve to guide the eye. The shaded region corresponds to a converted deposition temperature range of T

between 282 and 421 K using Eq. (1), a deposition rate of 0.3 µm/s, and a surface diffusion barrier of Q = 0.9 eV.

Fig. 11. Film configurations obtained at a constant deposition rate near 3 nm/ns and different combinations of temperature and vapor ratio of (a) T = 1000 K, Te:Cd = 0.8; (b) T = 1000 K, Te:Cd = 1.2; (c) T = 1200 K, Te:Cd = 0.8; and (d) T = 1200 K, Te:Cd = 1.2, with diatomic Te2 vapor. $\xi$: crystallinity; X: stoichiometry: $f_{Cd@Te}$ and $f_{Te@Cd}$: Cd@Te and Te@Cd antisite concentrations.

Fig. 12. Effect of vapor phase Te:Cd ratio on film crystallinity and stoichiometry. The lines only serve to guide the eye.

Fig. 13. Effect of vapor phase Te:Cd ratio on film antisite concentrations.

Fig. 14. Effect of vapor phase Te:Cd ratio on sticking coefficient. The lines only serve to guide the eye.

Fig. 15. Evolution of film configurations when deposited on a (10%) larger substrate at a temperature of 1000 K and a deposition rate near 2.5 nm/ns with diatomic $Te_2$ vapor. (a) time = 0.012 ns; and (b) time = 0.120 ns.

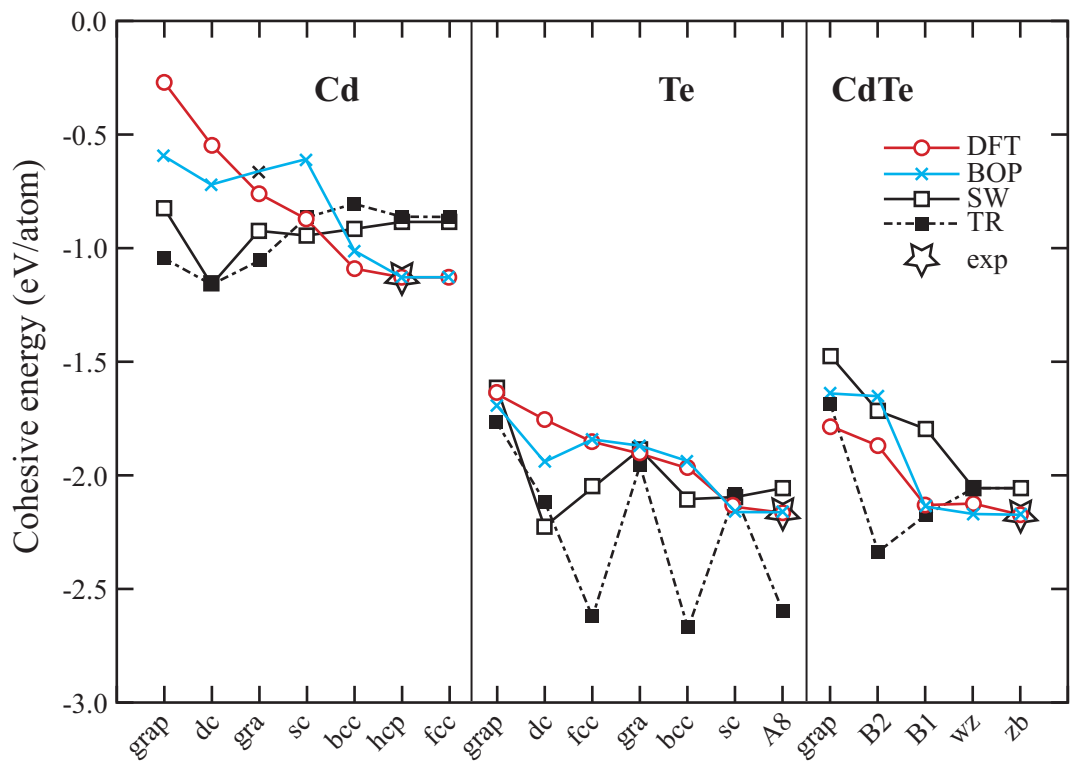

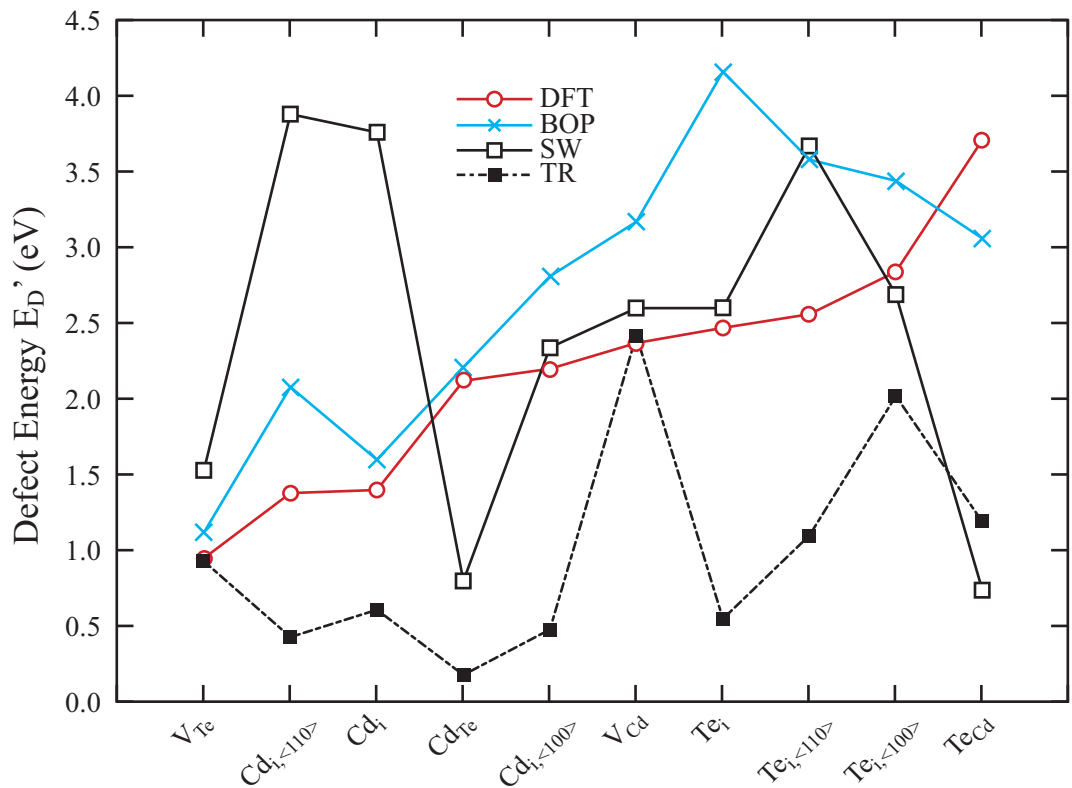

Figure 1

(a) R ~ 11.2 nm/ns at t = 0.3 ns

$\xi \sim 0.45$, $X \sim 1.01$, $f_{Cd@Te} = 0.075$, $f_{Te@Cd} = 0.092$

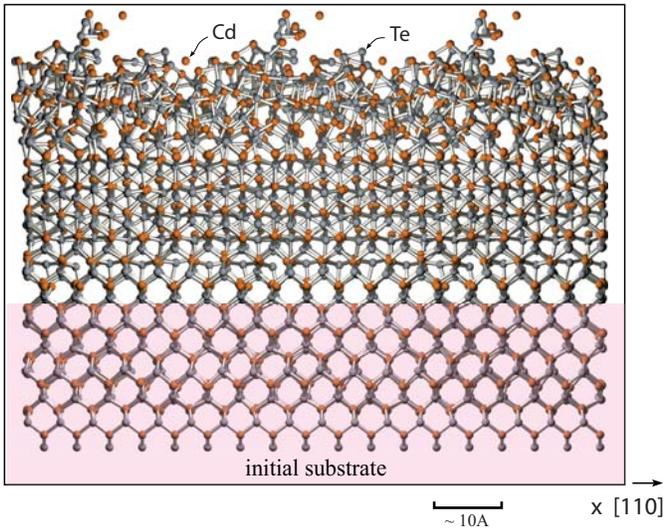

(b) R ~ 8.1 nm/ns at t = 0.4 ns

$E_i = 0.1$ eV
Te:Cd = 1
T = 1200 K
Te$_2$ vapor

$\xi \sim 0.69$, $X \sim 0.99$, $f_{Cd@Te} = 0.027$, $f_{Te@Cd} = 0.023$

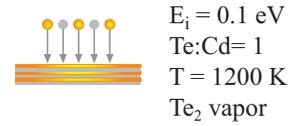
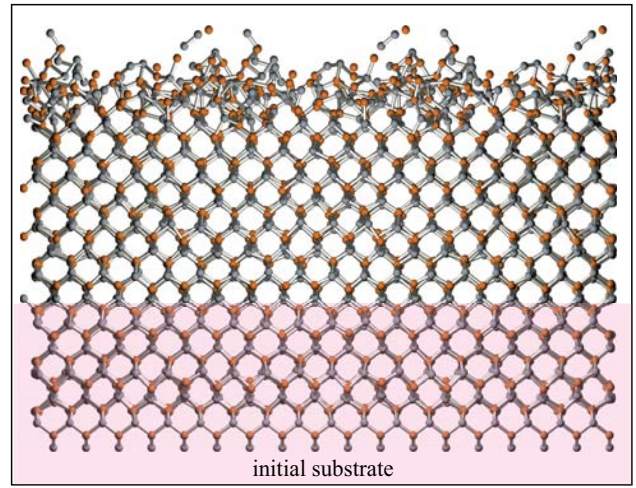

(c) R ~ 5.4 nm/ns at t = 0.6 ns

$\xi \sim 0.76$, $X \sim 1.02$, $f_{Cd@Te} = 0.012$, $f_{Te@Cd} = 0.017$

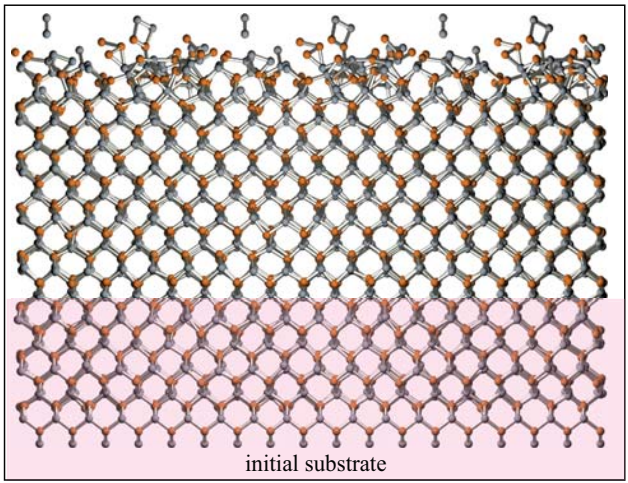

(d) R ~ 2.7 nm/ns at t = 1.2 ns

$\xi \sim 0.81$, $X \sim 1.02$, $f_{Cd@Te} = 0.002$, $f_{Te@Cd} = 0.008$

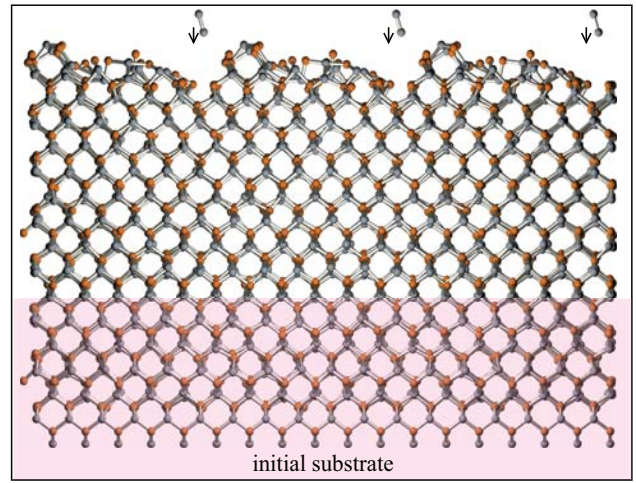

Figure 2

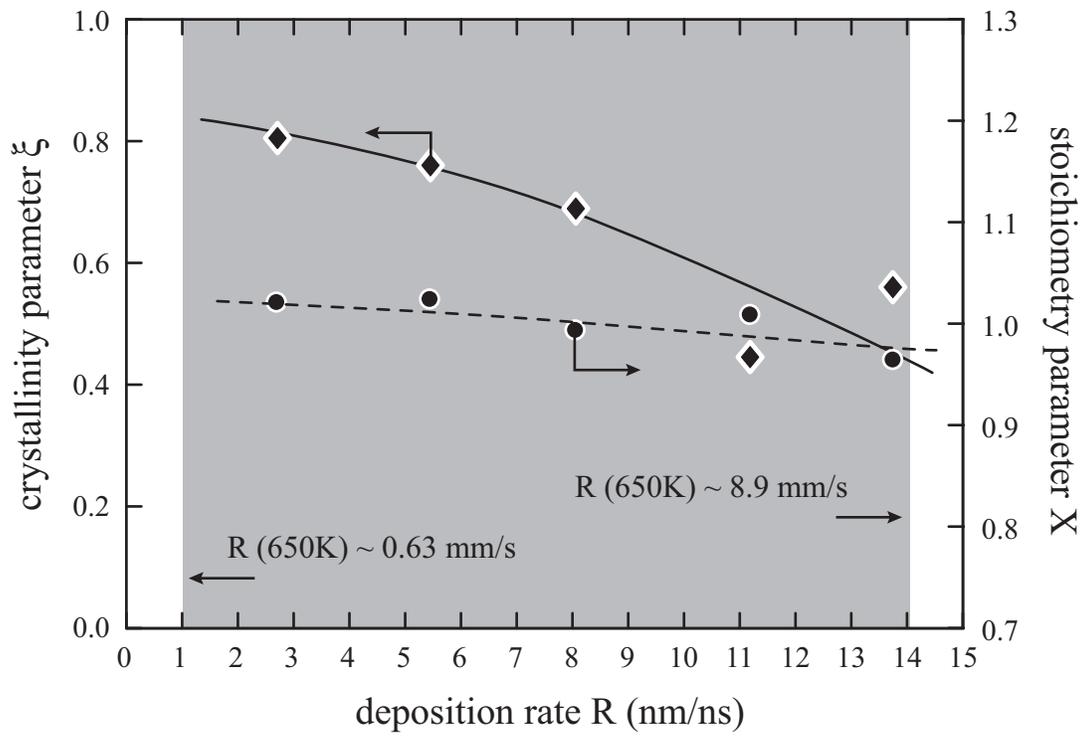

Figure 3

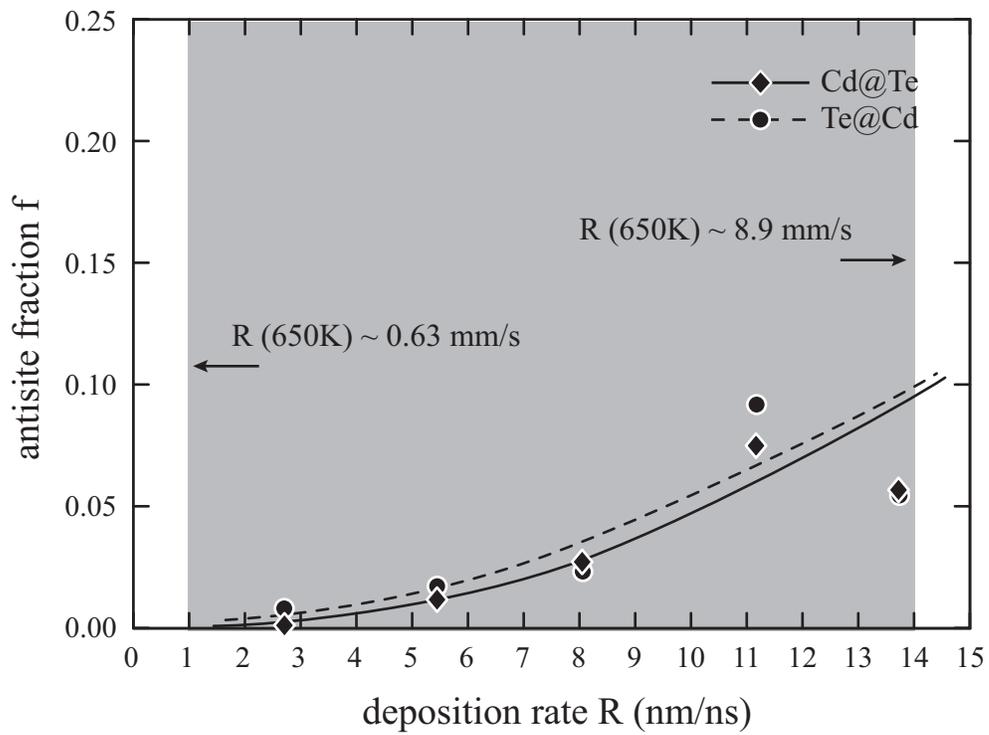

Figure 4

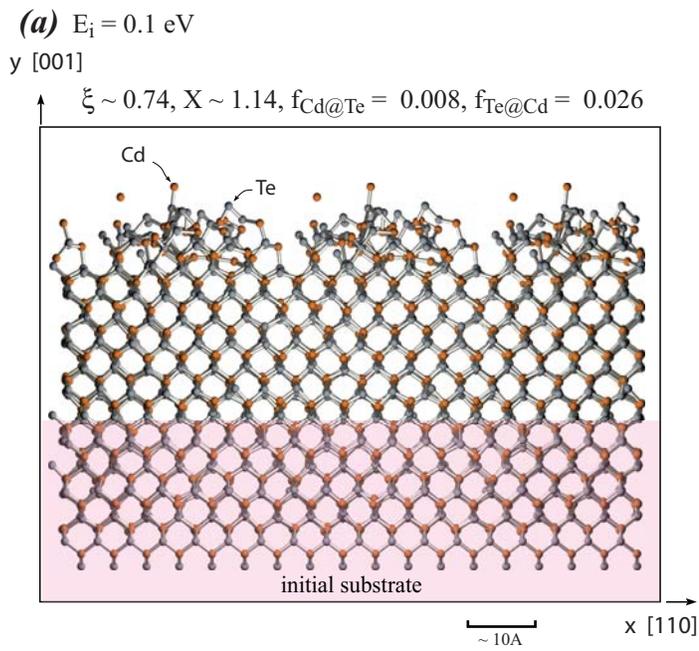 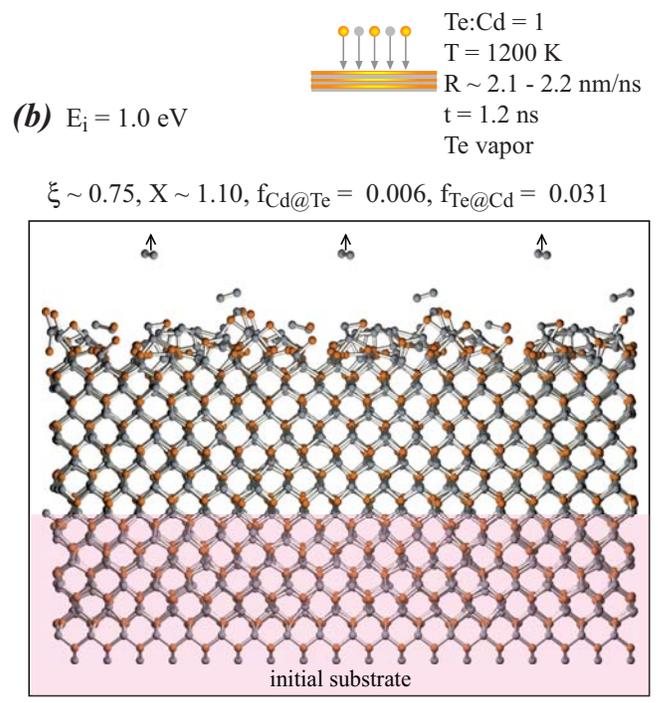

(a) $E_i = 0.1$ eV
y [001]
$\xi \sim 0.74$, $X \sim 1.14$, $f_{Cd@Te} = 0.008$, $f_{Te@Cd} = 0.026$

(b) $E_i = 1.0$ eV
$\xi \sim 0.75$, $X \sim 1.10$, $f_{Cd@Te} = 0.006$, $f_{Te@Cd} = 0.031$

Te:Cd = 1
T = 1200 K
R ~ 2.1 - 2.2 nm/ns
t = 1.2 ns
Te vapor

initial substrate

~ 10A   x [110]

Figure 5

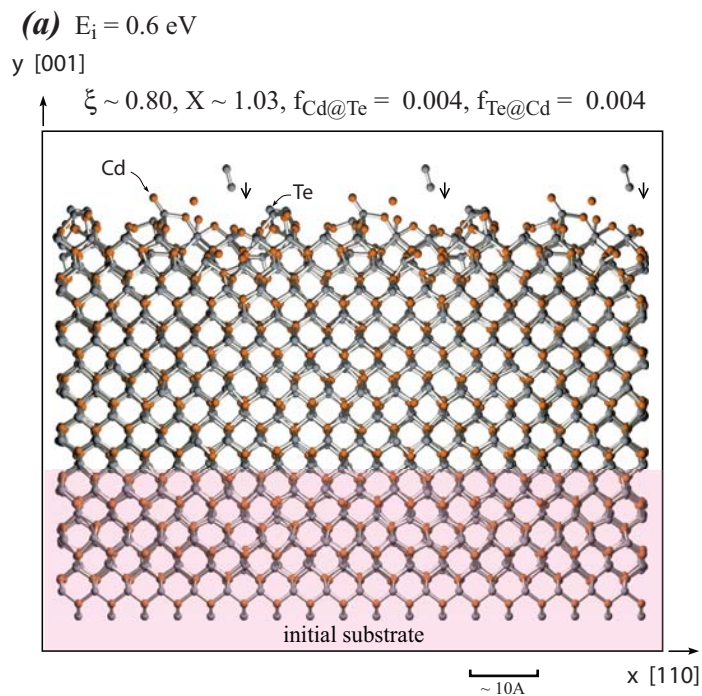 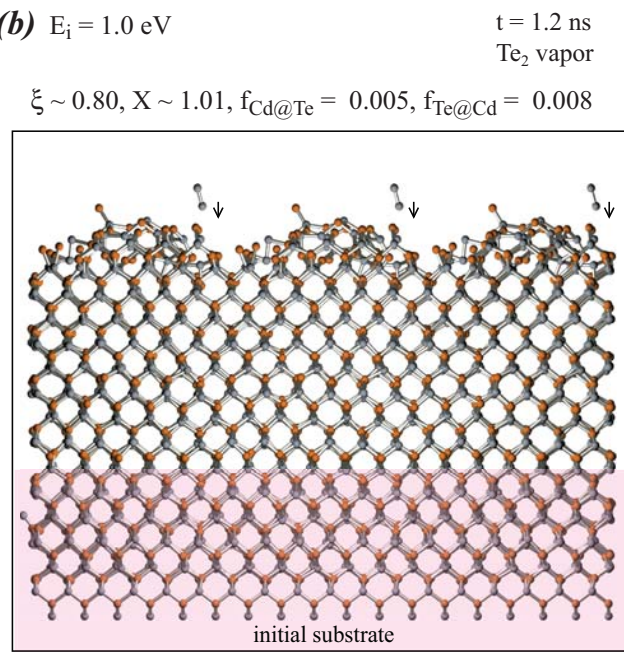

**(a)** $E_i = 0.6$ eV
$\xi \sim 0.80$, $X \sim 1.03$, $f_{Cd@Te} = 0.004$, $f_{Te@Cd} = 0.004$

**(b)** $E_i = 1.0$ eV
$\xi \sim 0.80$, $X \sim 1.01$, $f_{Cd@Te} = 0.005$, $f_{Te@Cd} = 0.008$

Te:Cd = 1
T = 1200 K
R ~ 2.7 nm/ns
t = 1.2 ns
Te$_2$ vapor

Figure 6

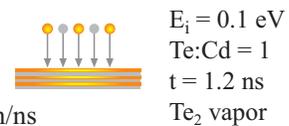

$E_i = 0.1$ eV
Te:Cd = 1
t = 1.2 ns
Te$_2$ vapor

**(a)** T = 700 K and R ~ 3.0 nm/ns

$\xi \sim 0.28$, X ~ 0.94, $f_{Cd@Te} = 0.149$, $f_{Te@Cd} = 0.155$

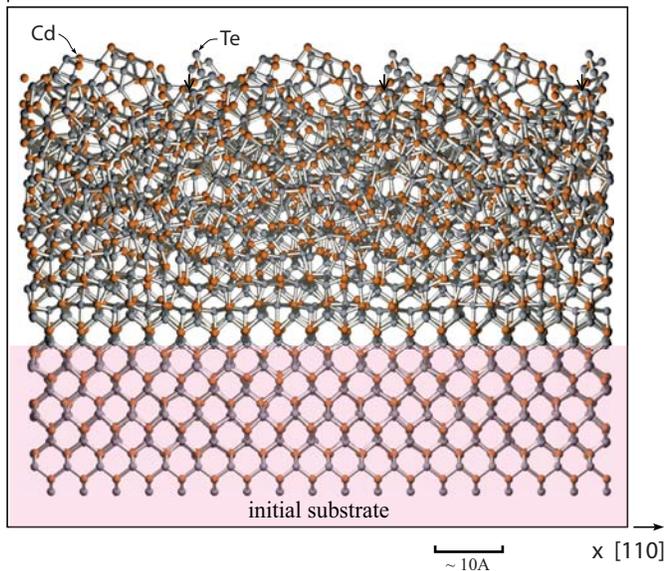

**(b)** T = 900 K and R ~ 2.9 nm/ns

$\xi \sim 0.79$, X ~ 0.99, $f_{Cd@Te} = 0.011$, $f_{Te@Cd} = 0.011$

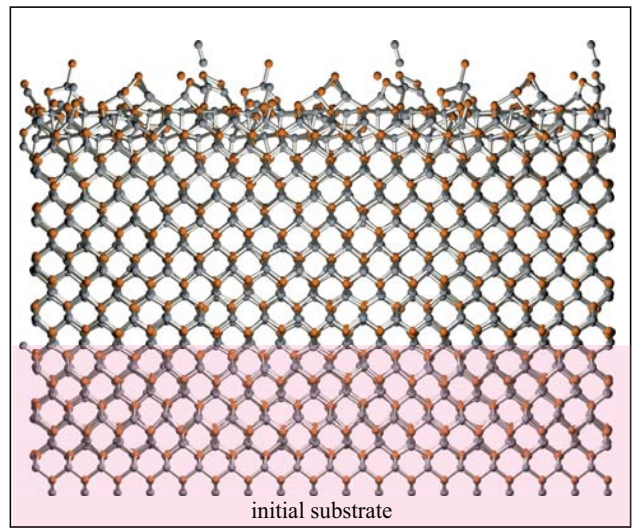

**(c)** T = 1000 K and R ~ 2.9 nm/ns

$\xi \sim 0.81$, X ~ 1.01, $f_{Cd@Te} = 0.007$, $f_{Te@Cd} = 0.009$

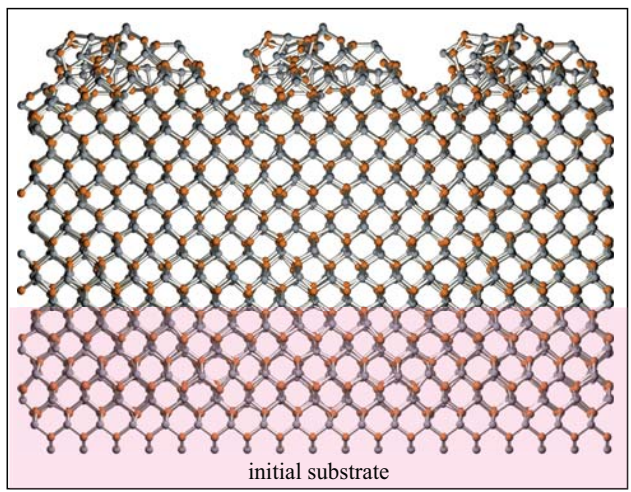

**(d)** T = 1100 K and R ~ 2.7 nm/ns

$\xi \sim 0.81$, X ~ 1.00, $f_{Cd@Te} = 0.002$, $f_{Te@Cd} = 0.002$

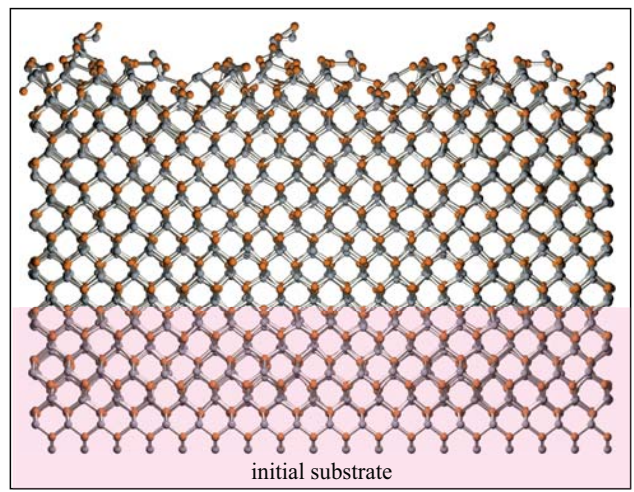

Figure 7

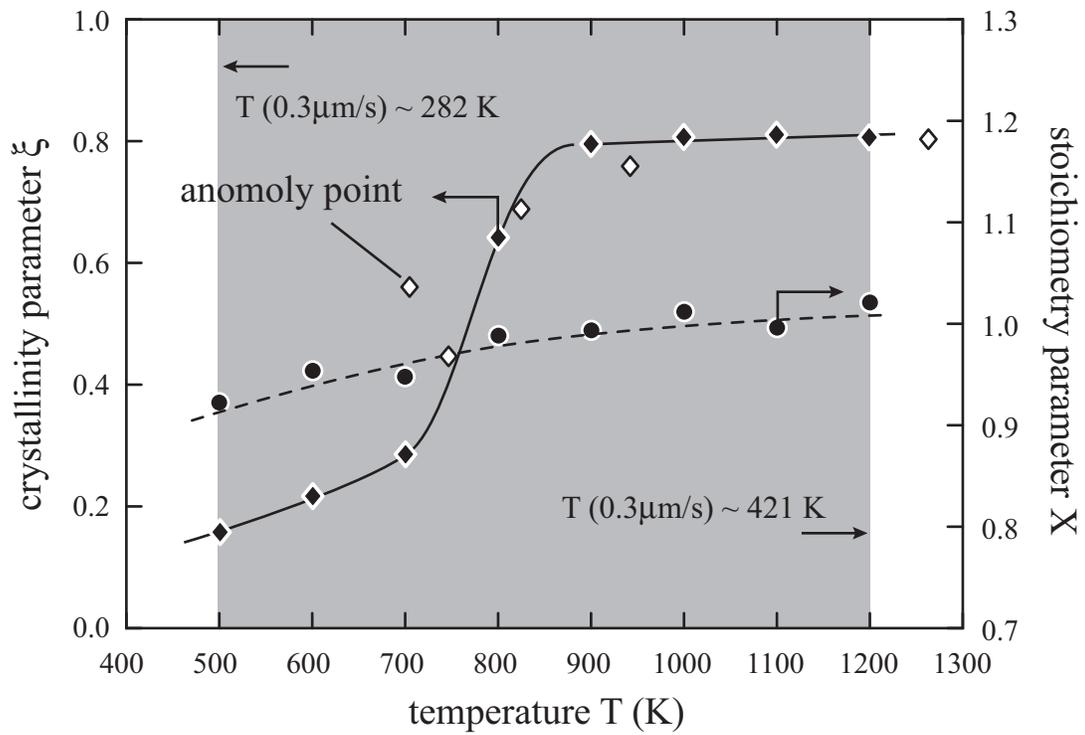

Figure 8

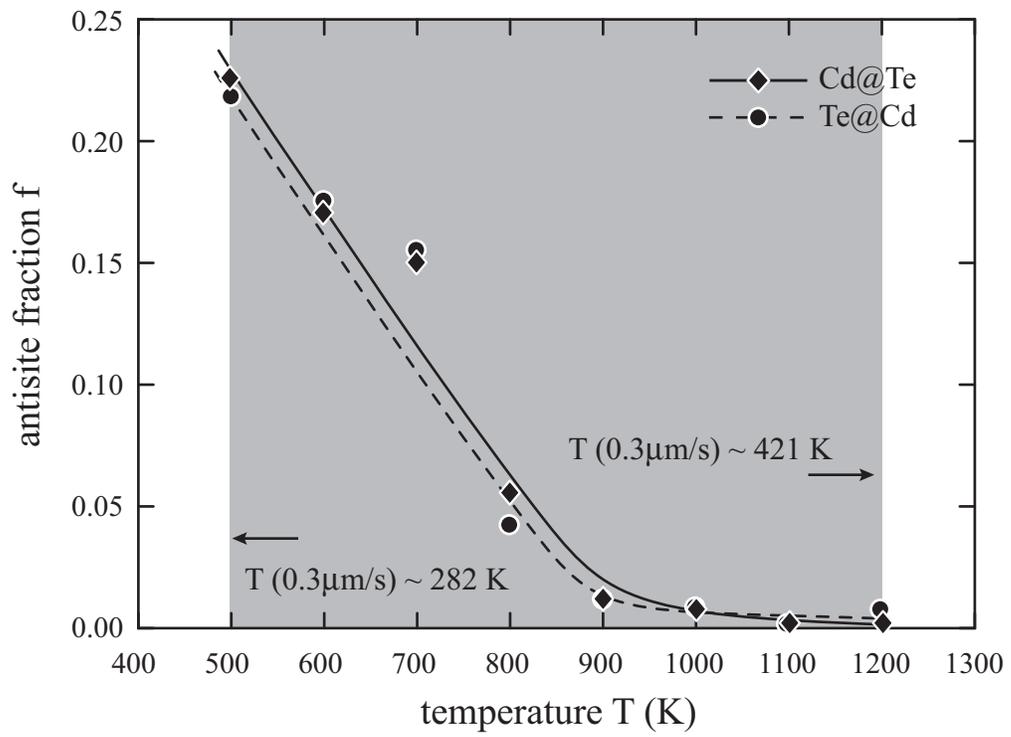

Figure 9

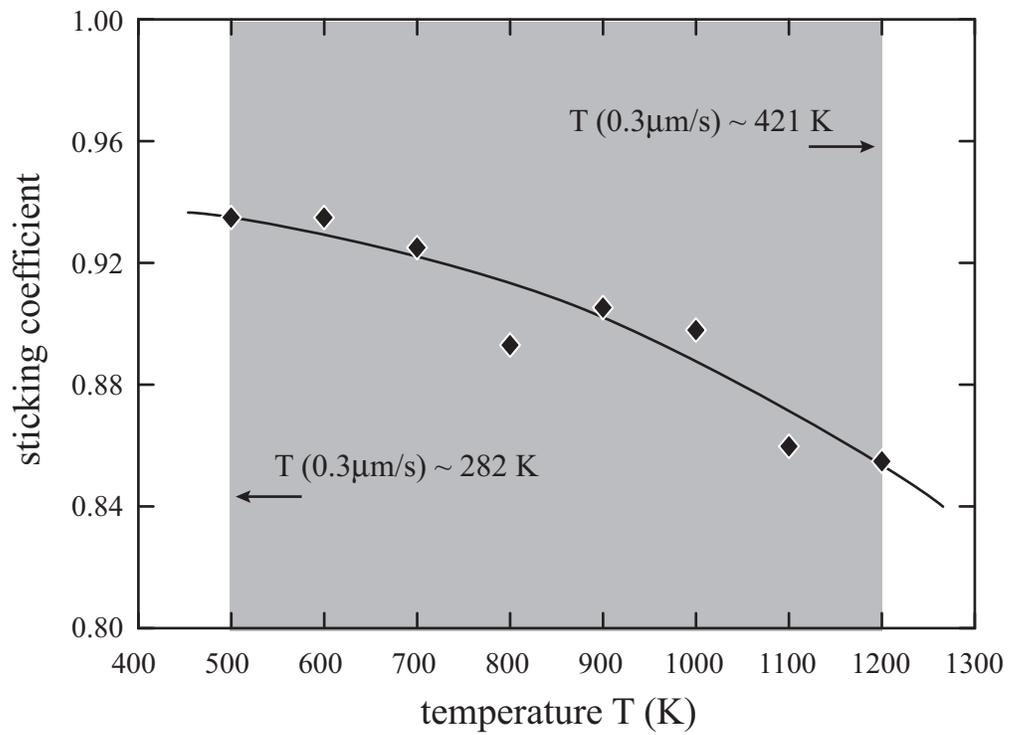

Figure 10

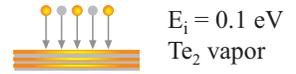

$E_i = 0.1$ eV
$Te_2$ vapor

**(a)** T = 1000 K, Te:Cd = 0.8, and R ~ 2.5 nm/ns

y [001]

$\xi \sim 0.78$, $X \sim 0.95$, $f_{Cd@Te} = 0.005$, $f_{Te@Cd} = 0.005$

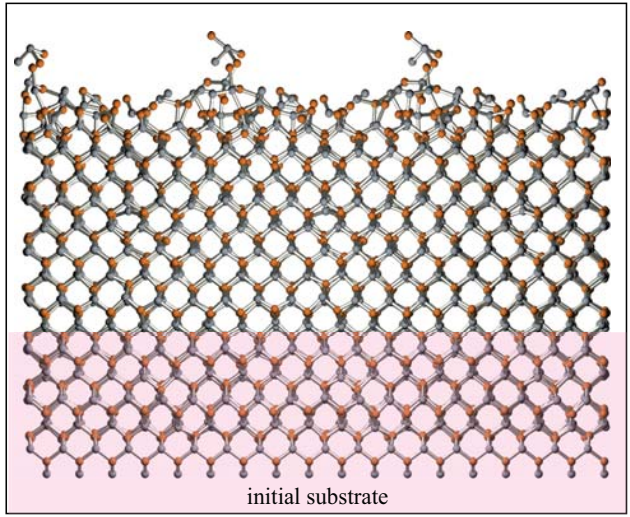

initial substrate

**(b)** T = 1000 K, Te:Cd = 1.2, and R ~ 3.08 nm/ns

$\xi \sim 0.69$, $X \sim 1.11$, $f_{Cd@Te} = 0.032$, $f_{Te@Cd} = 0.045$

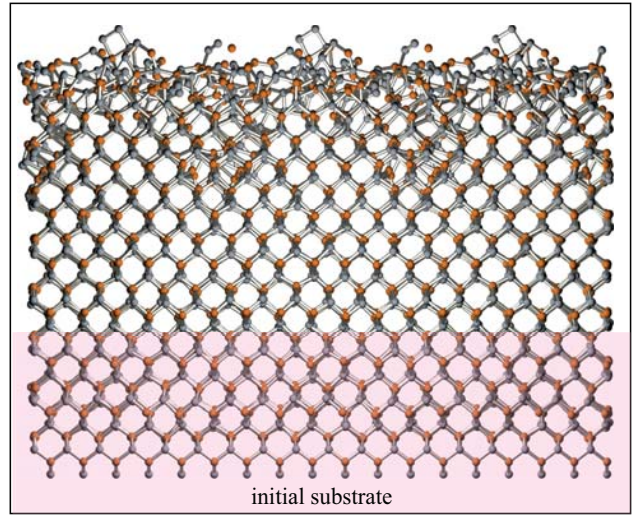

initial substrate

~ 10A   x [110]

**(c)** T = 1200 K, Te:Cd = 0.8, and R ~ 2.4 nm/ns

$\xi \sim 0.75$, $X \sim 0.98$, $f_{Cd@Te} = 0.015$, $f_{Te@Cd} = 0.015$

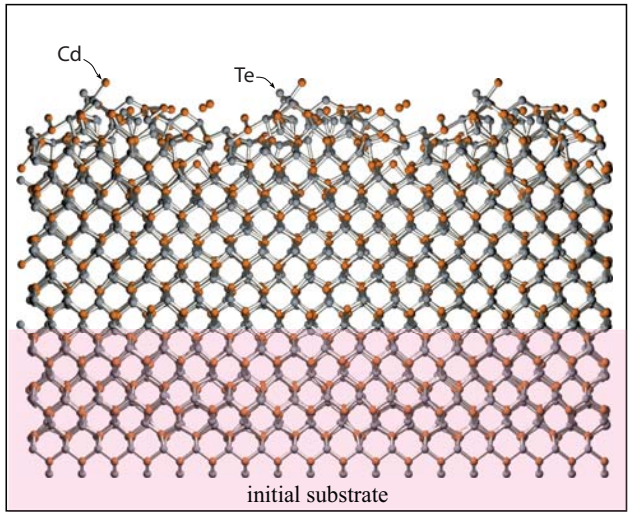

Cd    Te

initial substrate

**(d)** T = 1200 K, Te:Cd = 1.2, and R ~ 2.9 nm/ns

$\xi \sim 0.79$, $X \sim 1.09$, $f_{Cd@Te} = 0.003$, $f_{Te@Cd} = 0.021$

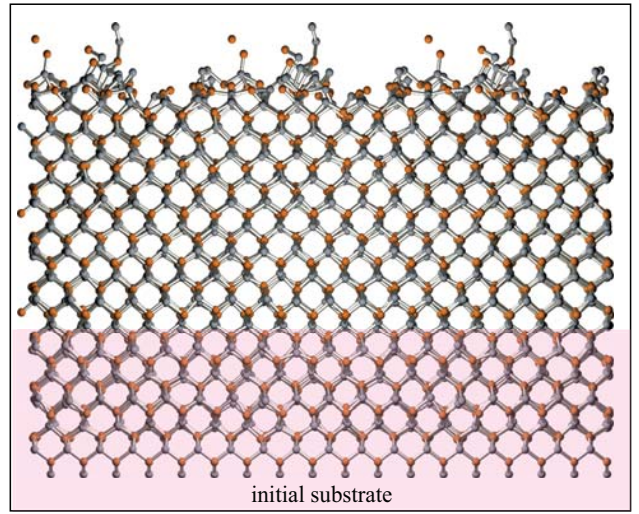

initial substrate

Figure 11

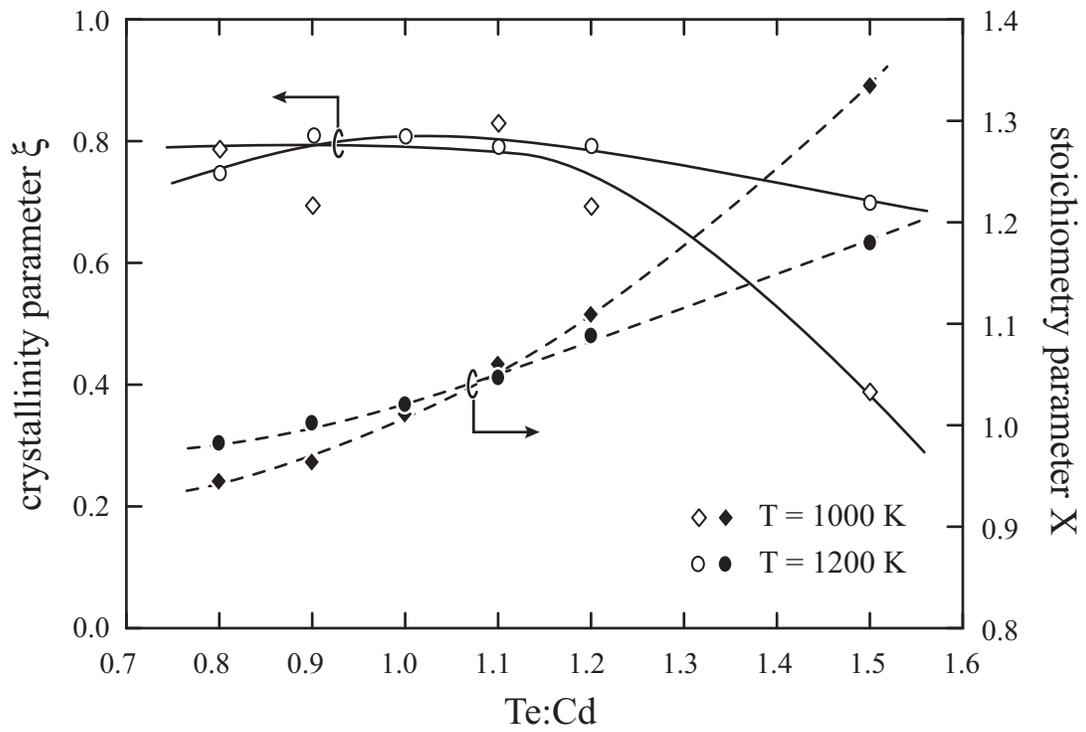

Figure 12

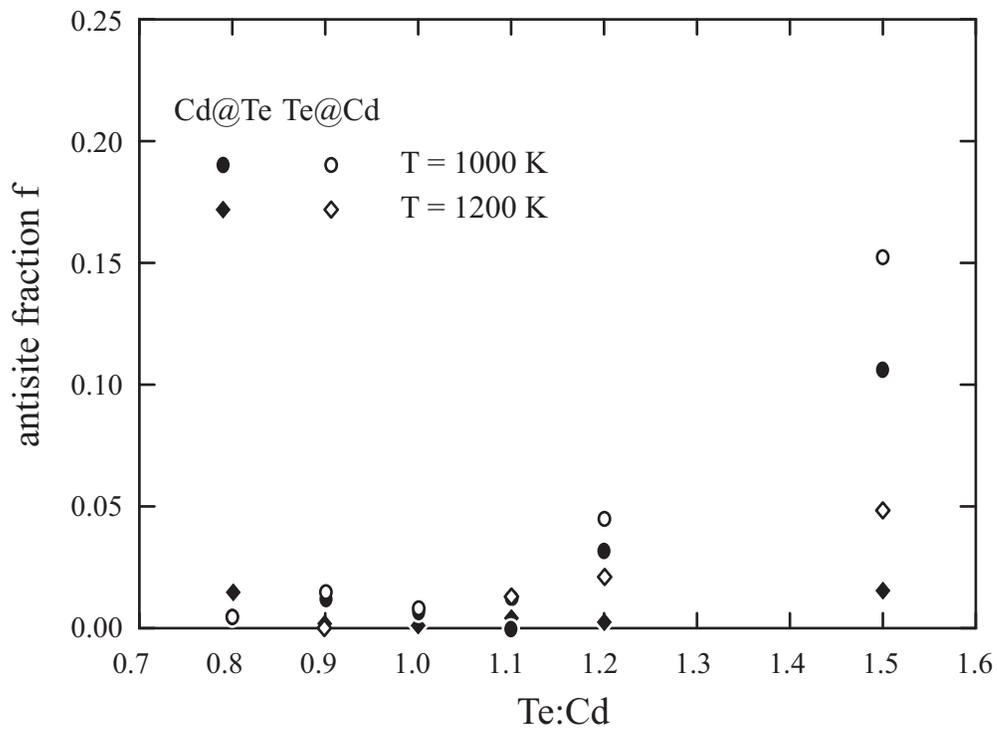

Figure 13

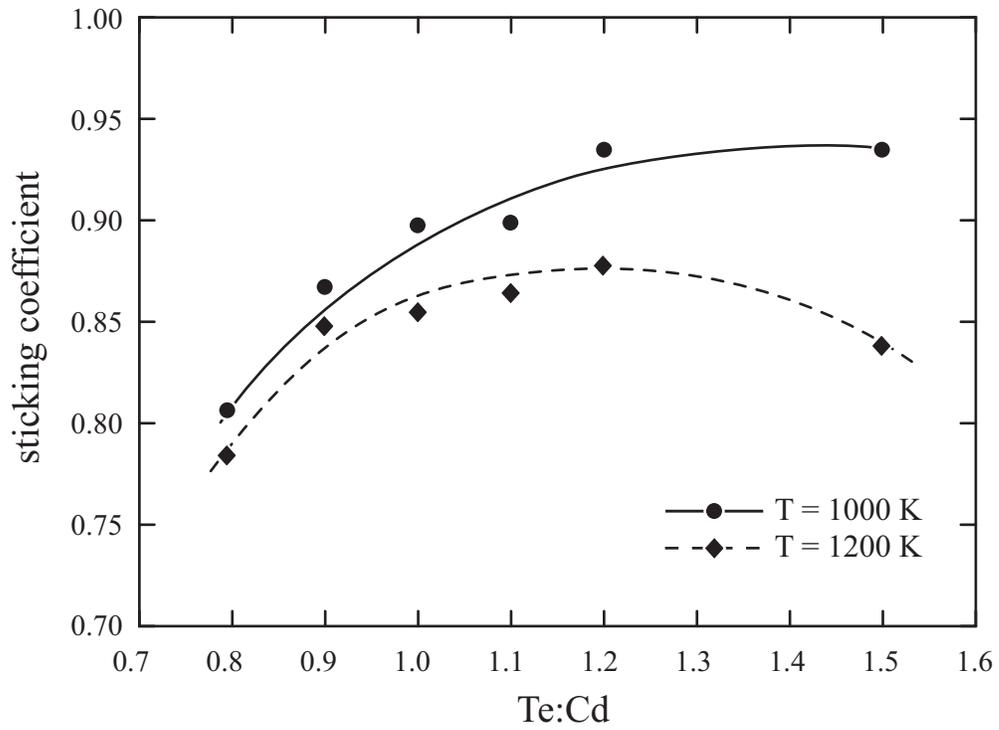

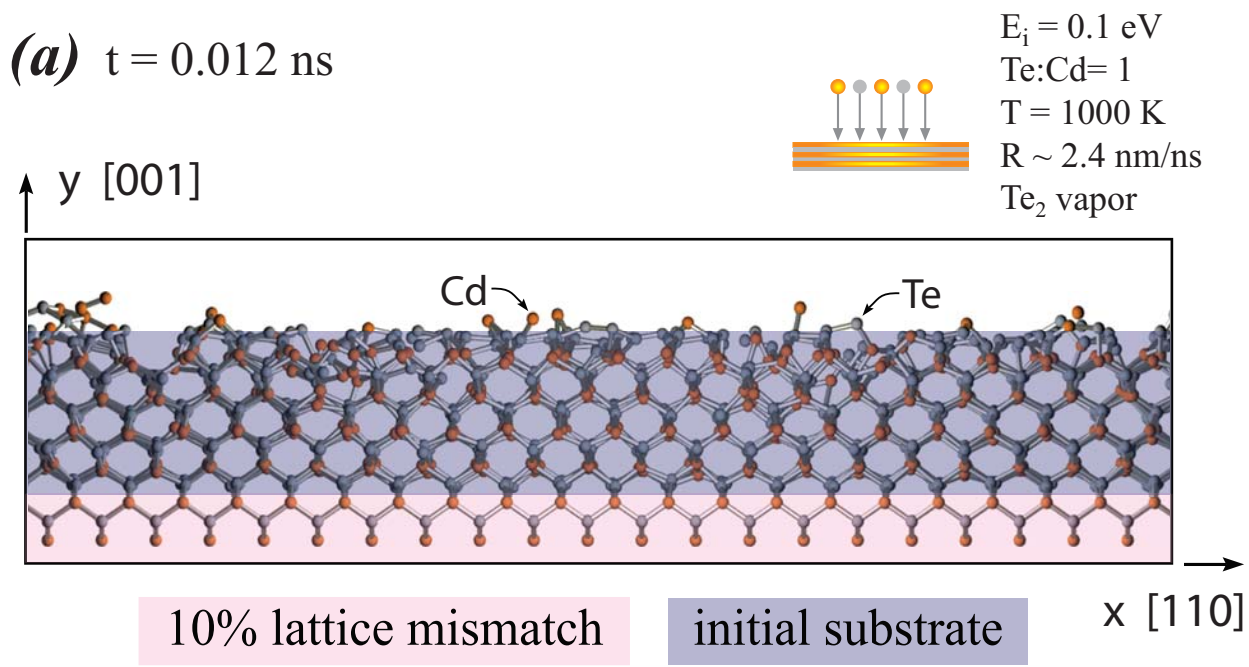
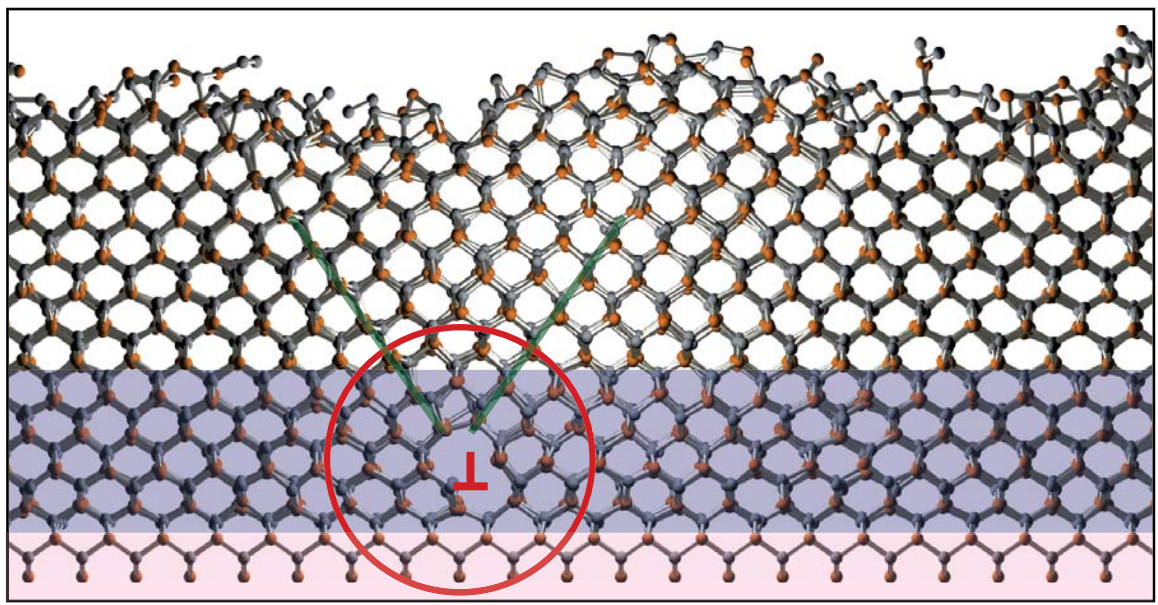

Figure 15